\documentstyle[12pt]{article}
\begin{document}

{\large
\begin{center}
{\bf      On the oscillation spectra of ultra compact stars  }
\end{center}}

\begin{center}
     Yasufumi Kojima $^1$, Nils Andersson $^2$ and Kostas D. Kokkotas $^3$
\end{center}
\begin{center}
{ \it  $^1$ Department of Physics, Tokyo Metropolitan University, \\
         Hachioji, Tokyo 192-03, Japan              }

{ \it $^2$ Department of Physics and Astronomy, \\
University of Wales College of Cardiff, PO Box 913,
Cardiff CF2 3YB, United Kingdom }

{ \it $^3$ Department of Physics, Aristotle University of Thessaloniki,
Thessaloniki 54006, Greece}

\end{center}
\vspace{0.5cm}

\begin{center}{  \bf  ABSTRACT   } \end{center}

Quasinormal modes of  ultra compact stars with uniform energy density
have been calculated.  For less compact stars, there is only one very
slowly damped polar mode (corresponding to the Kelvin f-mode) for each
spherical harmonic index  $l$.  Further long-lived modes become
possible for a sufficiently compact star (roughly when $M/R \ge 1/3$).
We compare the characteristic frequencies of these resonant polar modes
to the axial modes first found by Chandrasekhar and Ferrari [{\em
Proc.  Roy. Soc. London A} {\bf 434} 449 (1991)]. We find that the two
spectra approach each other as the star is made more compact. The
oscillation frequencies of the corresponding polar and axial modes
agree to within a percent for stars more compact than $M/R = 0.42$. At
the same time, the damping times are slightly different. The results
illustrate that there is no real difference between the origin of these
axial and polar modes: They are essentially spacetime modes.

\newpage

\begin{center}{ \bf   1. Introduction         }\end{center}

In nonradial pulsations of a spherically symmetric star, the polar and
the axial oscillation modes are quite different.  Most importantly, the
former couple to the density and pressure perturbations, while the
latter do not. This means that an incident gravitational wave can only
induce polar pulsation modes in the Newtonian limit.  However,
Chandrasekhar and Ferrari (1991b) have recently shown that this does
not remain true in the relativistic case.  Especially not if the
surface of the star is located inside the Regge-Wheeler potential peak;
when $ R$ is smaller than $ 3 M$ or so. In such a case, gravitational
waves can be
temporarily trapped inside the barrier, and the system exhibits damped
resonant oscillations also for axial perturbations.  Chandrasekhar and
Ferrari calculated eigenfrequencies for such resonant axial modes of
highly compact stars, $ M/R > 0.41,  $ where $ M $ and $ R $ are  the
mass and radius of the star (in geometrized units). They concluded that
only a few modes were possible for each stellar model, but that the
number of possible modes increased as the star became more compact.
The extensive (and numerically reliable) calculations by Kokkotas
(1994) do, however, make the existence of an infinite number of axial
modes for each compact stellar model seem very likely.  Most of these
modes are rather rapidly damped and could not be distinguished using
the resonance technique of Chandrasekhar and Ferrari.

In the stellar problem, axial modes depend only on the dynamical degree
of freedom associated with gravitational waves. Polar perturbations, on
the other hand, couple to the fluid oscillations and one would expect
the polar and the axial spectra to be quite different. For example, the
Newtonian p-modes adopt a small imaginary part (to account for
radiation damping) when  relativistic effects are included in the
analysis. These polar oscillation modes can have no analogue among the
axial modes. The situation is not at all that clear for the highly
damped (w) modes (Kojima 1988, Kokkotas and Schutz 1992, Leins {\em et
al.} 1993 and Andersson {\em et al.} 1995a). These  are mainly
spacetime modes, and there is no apparent reason why similar axial
modes should not exist.

The situation prompts some interesting questions:  Are there resonant
polar modes analogous to the axial modes found by Chandrasekhar and
Ferrari  (1991b)? What is the relation between the branch of  highly
damped axial modes (cf. Kokkotas 1994) and the polar w-modes?  Do the
characteristic frequencies of the axial and the polar modes approach
each other as the star is made more and more compact?  One would
certainly expect an affirmative answer to the first of these
questions.  There is no reason why such polar modes should not exist.
In fact, one would expect the spectrum of polar modes to approach that
of the axial modes for very compact stars. This expectation is based on
the experience from studies of gravitational perturbations of a
Schwarzschild black hole. In that case, the polar and  axial
perturbations are related by a certain transformation, and the
corresponding quasinormal-mode frequencies are identical (Chandrasekhar
1983).  In the case of stars, the equations that govern the
perturbations in the exterior vacuum are identical to those for a black
hole.  It therefore seems likely that the two stellar spectra approach
each other (in some sense) as the star becomes more compact, {\em
i.e.}, as more of the black-hole potential barriers come into play.
Moreover, an assumption that the branch of highly damped axial modes
(Kokkotas 1994) corresponds directly to the polar w-modes -- that have
so far only been studied for polytropic stellar models -- is
straightforwardly tested.  If that is the case, highly damped axial
modes should exist also for less compact stars ($M/R \ge 1/3$), even
though such models cannot support ``trapped'' modes with a very slow
damping. If these suggestions can be proved true it would illustrate
beyond doubt that the axial modes -- as well as the polar w-modes --
are in all essential respects ``spacetime'' modes, the properties of
which are determined only by the curvature of spacetime.

In this short paper we examine compact stellar models with uniform
energy density. Although astrophysically unrealistic, such models have
the advantage that we need not worry about fluid oscillations.
Moreover, these models can easily be made very compact. The present
discussion concerns stellar models that approach the limit of
compactness imposed by General Relativity: $ M/R \le 4/9$. It should be
remembered that stars as compact as that, almost certainly, do not
exist in our universe. A useful comparison is provided by the values
$R= 10\ {\rm km}$ and $M=1.4M_\odot$ often used in rough calculations
involving neutron stars. These values correspond to $M/R \approx 0.21$.
Nevertheless, the ultracompact uniform density star serves as a
reasonably simple model problem that can help us understand better the
origin of the various oscillation modes of relativistic stars.


\vspace{ 0.5cm }
\begin{center}{ \bf   2. Comments on stars with uniform energy density   }
\end{center}

We assume that the energy density is constant throughout the star in
its equilibrium state. We also assume that the Eulerian change of the
energy density vanishes. In Newtonian pulsation theory, this
approximation leads to a single oscillation mode; the Kelvin f-mode.
In fact, we use the uniform density approximation to avoid
``uninteresting'' fluid oscillations that give rise to the p- and
g-modes. These are well understood and there is no reason why  we
should include them here.

The equations that govern axial and polar perturbations of uniform
density stars are easily derived from the equations used for more
realistic stellar models.  Consequently, we will not give many details
here. Rather, we will outline the approach that we have used in each
case, and refer the reader to the original papers for more details.

In the case of polar perturbations, the relevant equations can be
obtained by imposing  $ \delta \varrho =0  , $ or equivalently $ C^{-2}
=0 $ in eqs. (38)--(39) of Kojima (1992). Explicit expressions for the
remaining variables (such as the pressure) for a uniform density star
has been given by Chandrasekhar and Ferrari (1991b).  The basic
equations inside the star become two second-order differential
equations for certain components of the metric perturbations ($ H_{0,l}
$ and $ K_{l}$).  Physical solutions to these two coupled equations
must be regular at the centre of the star. One must also ensure that
the Lagrangian change in pressure vanishes at the stellar surface.

Outside the star -- in vacuum -- the perturbation equations  reduce to
a single second-order differential equation. In the case of polar
perturbations, this is the Zerillli equation familiar from studies of
Schwarzschild black holes (Chandrasekhar 1983). A physically acceptable
solution to the problem for the stellar interior generally corresponds
to a linear combination of outgoing and incoming waves at infinity;
\begin{equation}
    K_{l}   \longrightarrow
      A_{\rm out} \exp ( -i \sigma r^* ) +  A_{\rm in}  \exp (i \sigma
   r^* )  , \quad {\rm as}\ r^\ast \to +\infty \ ,
\label{asympt}\end{equation}
where  $ r^* $ denotes the standard tortoise coordinate. $ A_{\rm in}
$  and  $ A_{\rm out} $ are the constant amplitudes of the incoming and
outgoing gravitational waves at infinity, respectively. We assume that
all perturbations have time-dependence $\exp (i\sigma t)$. Quasinormal
modes of the stellar system are distinguished by purely outgoing waves
at spatial infinity, {\em i.e.}, $ A_{\rm in} =0  .$ This condition is
satisfied only for a discrete set of  complex frequencies.

We use the resonance method developed by Chandrasekhar and Ferrari
(1991a) to determine the slowest damped polar modes for the uniform
density model. Assuming that the  imaginary part ($\sigma_I$) of the
eigenvalue is small, we calculate  $ | A_{\rm in}  | $ for real values
of $ \sigma.  $ The curve of $ | A_{\rm in}  |  $  then exhibits a deep
minimum as $| A_{\rm in} |^2  =  {\rm const} \times \{ ( \sigma -\sigma
_R )^2 + \sigma _I ^2 \}. $
{}From the position and shape of such minima it is straightforward to
deduce $ \sigma _R $ and $ \sigma _I , $ which correspond to the real
and imaginary parts of our complex eigenfrequencies.

The resonance method can only be trusted for modes which are slowly
damped. For highly damped modes it must be replaced by an iterative
method. Such a method must be able to deal with the problem that the
quasinormal-mode eigenfunctions diverge at spatial infinity. (It is
clear from (\ref{asympt}) that $K_{l}$ diverges when the imaginary part
of $\sigma$ is positive.) Methods that have been developed to handle
this difficulty include the WKB method used by Kokkotas (1994) and the
numerical integration scheme of Andersson {\em et al.} (1994a).
However, in the present analysis we do not wish to study highly damped
modes. We are primarily interested in the ``trapped'' modes that occur
as the star becomes extremely compact. These are going to be slowly
damped so we are not much restricted by the limitations of the
resonance method.

In the case of axial perturbations the problem has been described in
detail by Chandrasekhar and Ferrari (1991b). Since there is no coupling
to the fluid, the interior problem can be formulated as a single
second-order differential equation analogous to the equation for the
exterior vacuum; the Regge-Wheeler equation (Chandrasekhar 1983). As in
the polar case, the physically acceptable solution -- that is regular
at the centre of the star -- generally corresponds to a combination of
out- and ingoing waves at infinity, and axial quasinormal modes can be
computed in exactly the same way as the polar modes. In our
calculations for axial modes we use the approach of Andersson {\em et
al.} (1995b). This scheme was specially developed for highly damped
modes (it finds the modes by iteration), but it works equally well for
slowly damped modes.

The fact that the interior problem can be formulated as a single
Schr\"odinger-like differential equation for axial perturbations
inspired Chandrasekhar and Ferrari (1991b) to the discovery of slowly
damped axial modes. The general idea is that, when the surface of the
star lies inside the peak of the Regge-Wheeler potential barrier, the
system can support ``quasi-bound'' states that slowly leak out through
the barrier to spatial infinity. That similar modes should exist for
polar perturbations is not as easily made apparent.

\vspace{ 0.5cm }
\begin{center}{ \bf  3. Discussion of numerical results       }\end{center}

We have calculated the slowest damped quasinormal-mode frequencies for
both axial and polar perturbations of uniform density models with a
varying degree of compactness. All calculations discussed here are for
quadrupole modes ($l=2$).  In Figure 1 we show the amplitude of
incoming waves at infinity  ($ | A_{\rm in}  | $)  as a function of
real frequency $ \sigma $ for four stellar models. The figure adheres
to polar modes and  contains information relevant for mode-calculation
by the resonance method.  One resonance minimum is evident in the
stellar model with $ M/R =0.2 .$ The oscillation frequency is $ \sigma
_R (R^3/M)^{1/2} = 0.887$ and the damping rate is governed by  $ \sigma
_I (R^3/M)^{1/2} = 3.6 \times 10^{-4}$. We have verified that this mode
tends to the frequency of the Kelvin f-mode; $ \sigma _R (R^3/M)^{1/2}
= \sqrt{ 2l (l-1)/(2l+1) } =0.894 $ (cf. Tassoul 1978) in the Newtonian
limit (as $ M/R  \rightarrow 0 $) . It is clear that, there is only one
really slowly damped oscillation mode in less compact stars. This is
expected from Newtonian pulsation theory. We know, of course, that
there will also be an infinite number of highly damped (w) modes, but
as the star gets less relativistic the damping of these modes increases
(Kokkotas \& Schutz 1992). Hence, such modes cannot be identified by
the resonance method.

\vspace{0.5cm} \begin{center} \begin{tabular}{|c|}
 \hline   Figure  1 near here  \\    \hline
\end{tabular}  \end{center} \vspace{0.5cm}

For the model with $ M/R=0.40 $ one can see a tiny dent at  $ \sigma
(R^3/M)^{1/2} \sim 1.7 $ in Figure 1. The resonance minimum becomes
evident in the model with $ M/R=0.42 , $ although the frequency has
then shifted to $ \sigma  (R^3/M)^{1/2} \sim 1.3. $ It is possible to
determine this mode numerically for $ M/R > 0.4. $ This criterion is
almost the same as that obtained in the axial-mode study of
Chandrasekhar and Ferrari (1991b). The resonance point near $ \sigma
(R^3/M)^{1/2} \sim 1.7 $ in the model with $ M/R=0.40  $ is so
ambiguous that reliable calculation by the resonance method is
impossible.  The need for an alternative approach, for this and less
compact models, is clear. As the star becomes more compact, further
resonant modes become evident -- as shown in the model with $ M/R=0.44$
in Figure 1.  The number of  distinguishable resonances clearly
increases with the compactness.

Kokkotas (1994) calculated many higher overtone axial modes. His
results have recently been confirmed as reliable by Andersson {\em et
al.} (1995b).  The imaginary part of most of those  modes is so large
that the resonance method can not be used to calculate them, however.
But it is important to remember that each stellar model supports an
(almost certainly) infinite number of axial modes.

In Figure 2 we compare the quasinormal-mode frequencies for polar
perturbations to those for axial perturbations. The real and imaginary
part of each mode-frequency are studied separately as a function of the
compactness of the star, $ M/R $. Our numerical calculation of polar
modes is limited to $ \sigma_I / \sigma_R \leq 0.01 $ and it should be
remembered that the minima of the amplitude  $ | A _{\rm in} | $ become
less clear as the star becomes less compact: The imaginary part of each
mode increases.  We have also limited the study to  $ M/R \leq 0.44 .$
This value is slightly smaller than the extreme case allowed by General
Relativity; $M/R=4/9 =0.444 \cdots . $ As the star becomes extremely
compact, there seems to be a mode-crossing between the f-mode and the
first spacetime mode. This occurs near $ M/R =0.44 $ in  Figure 2. When
this happens two minima of   $ | A_{\rm in}  | $ merge, and it is very
difficult to calculate eigenfrequencies near the crossing point with
the resonance method.

\vspace{0.5cm} \begin{center} \begin{tabular}{|c|}
 \hline   Figure  2 near here  \\    \hline
\end{tabular}  \end{center} \vspace{0.5cm}

As can be seen from Figure 2, the agreement between the polar and the
axial modes is  good for all the studied models. In fact, the physical
oscillation frequencies of the axial and polar modes agree to within a
percent for models more compact than $M/R = 0.42$. This close agreement
does, indeed, support the idea that the axial and polar modes approach
each other as the star gets very compact (when it,  in a loose sense,
``approaches''  a black hole). In fact, the discrepancy is at the same
level as the numerical uncertainties expected in the resonance method.
It is also apparent from Figure 2 that the  discrepancy between the
imaginary parts is, in general, larger. We believe that this is to some
extent due to numerical inaccuracies. When the imaginary part is many
orders of magnitude smaller than the corresponding real part it is
difficult to determine $\sigma_{I}$ with great accuracy.

\vspace{ 0.5cm }
\begin{center}{ \bf  4. Concluding remarks  }\end{center}

Outside a star, both polar and axial perturbations are described by a
second-order differential equation with an effective potential barrier,
the peak of which lies at roughly $r \approx 3M$.  When the stellar
surface lies inside the peak of this barrier, gravitational waves can
be temporarily trapped in a way that is reminiscent of ``quasibound''
resonance states in quantum scattering.  These trapped
gravitational-wave modes then decay gradually.  The modes studied here
correspond to this situation. As the star becomes more compact the
decay rate, {\em i.e.}, the imaginary part of the mode-frequency,
decreases because  the  potential ``well'' inside the barrier gets
deeper. Moreover, as the star is made increasingly compact more of the
black-hole potentials -- that govern the perturbation in the exterior
spacetime -- is unveiled. Since the axial and polar-mode spectra are
manifestly identical for Schwarzschild black holes, one might expect
the two spectra to approach each other for stars of increasing
compactness.  \footnote{It should, of course, be emphasized that the
spectra studied here are considerably different from that of a black
hole. This is, however, to be expected; There is simply no continuous
transformation from a stellar model to a black hole. The inner boundary
condition, {\em i.e.}, the existence of a horizon is but one crucial
factor that causes the differences.} The results obtained in the
present study support this view. It is clear that, there are
considerable similarities between the quasinormal modes for polar and
axial perturbations of compact stellar models. Since there are no fluid
motion in a uniform density star the oscillation modes studied here are
related to the dynamical degree of freedom of gravitational waves. With
exception of the f-mode, there are no real differences between the
mechanisms behind the polar and the axial modes here:  They are all
essentially ``spacetime'' modes.

In many ways, the modes studied here resemble the w-modes that has been
found for polar perturbations of polytropic stellar models (Kojima
1988, Kokkotas \& Schutz 1992, Leins {\em et al.} 1993, Andersson {\em
et al.} 1995a). For these modes, the gravitational-wave degrees of
freedom play an essential role.  All w-mode calculations to date have
been for much less relativistic stellar models than those considered
here. However, as can be inferred from Figure 2, the imaginary part of
each ``trapped'' mode increases (roughly) exponentially as $ M/R$
decreases. If we recall that the slowest damped w-mode for a polytropic
model with $M/R = 0.297$ is $(2.910+0.346i) (R^3/M)^{1/2}$ (Andersson
{\em et al.} 1995a), this makes it, indeed, plausible that the modes
studied here are intimately connected to the w-modes in a less compact
star.  If that is the case, one should be able to confirm the existence
of very long-lived w-modes for compact polytropes.

Furthermore, it seems logical to predict that highly damped axial modes
should exist for much less relativistic stars. Noone has actually
searched for such modes, but there is no apparent reason why they
should not exist. The argument has been that quasinormal modes rely
upon coupling to the fluid for their existence. We believe that this
argument is outdated, and that the results discussed here (and also the
recent ones by Kokkotas (1994) and Andersson {\em et al.} (1995b) )
demonstrate that these are modes which,  in all essential respects, are
due to the curvature of spacetime in the neighbourhood of the star.  To
verify the existence of highly damped axial modes for less relativistic
models, and test the correspondence between the axial and polar spectra
further, one must calculate eigenfrequencies with large imaginary
parts. That task is beyond the scope of the present work, but we aim to
address it in the future.

\vspace{ 0.5cm }
\begin{center} { \bf      Acknowledgements } \end{center}

This work was inspired by a discussion at the Seventh Marcel Grossman
meeting at Stanford in July 1994. During that discussion  Prof. S.
Chandrasekhar advanced the view that the axial and the polar spectra
would approach each other as a star is made more compact. At that time
we had preliminary results that supported this view, but the discussion
prompted us to study the problem in more detail.

This work was supported in part (Y.K.) by the Japanese Grant-in-Aid for
Scientific Research on Priority Area (03250212, 04234104, 04234209,
05218208). N.A. acknowledges support from SERC in the United Kingdom.

\vspace{ 0.5cm }
\begin{center}{ \bf    REFERENCES  }\end{center}

\noindent
Andersson, N., Kokkotas, K.D. and Schutz, B.F. 1995a {\em A new numerical
approach to the oscillation modes of relativistic stars} to appear in
{\em Mon. Not. R. Astron. Soc.}.

\noindent
Andersson, N., Kokkotas, K.D. and Schutz, B.F. 1995b {\em Spacetime
modes of relativistic stars} in preparation.

\noindent
Chandrasekhar, S. 1983
{\it The Mathematical Theory of Black Holes}
Oxford: Clarendon Press.

\noindent
Chandrasekhar, S.  and  Ferrari, V. 1991a  {\em Proc. R. Soc. Lond.
 A} {\bf 432, }  247-249.

\noindent
Chandrasekhar, S.  and Ferrari, V. 1991b  {\em Proc. R. Soc. Lond.
 A} {\bf 434, }  449-457.

\noindent
Kokkotas,  K.D.,  1994  {\em  Mon. Not. Roy. Astr. Soc} ,
 { \bf 268, } 1015-1018.

\noindent
Kokkotas , K.D., and Schutz B.F., 1986 {\em Gen. Rel. and Grav.} ,
  { \bf 18, } 913-921.

\noindent
Kokkotas , K.D., and Schutz B.F., 1992  {\em Mon. Not. Roy. Astr. Soc.},
  { \bf 255, } 119-128.

\noindent
Kojima,  Y.,  1988  {\em Prog. Theor. Phys. } { \bf 79, } 665-675.

\noindent
Kojima,  Y.,  1992 {\em Phys. Rev. D.} { \bf 46, } 4289-4303.

\noindent
Leins M., Nollert H-P., and Soffel M.H. 1993 {\em Phys. Rev. D}
{\bf 48} 3467-3472.

\noindent
Tassoul, J., 1978
{\it The theory of rotating stars }
Princeton: Princeton Univ. Press.

\newpage
\begin{center}{ \bf      Figure caption }\end{center}

\noindent
Fig.1.
The amplitude of the incoming gravitational wave at spatial infinity, $
| A_{\rm in}  | $, as a function of frequency,  $  \sigma (R^3/M)^{1/2}
$, for different uniform density stellar models. This figure is for
polar perturbations.
{}From the minima  quasinormal frequencies can be
calculated.

\vspace{ 0.5cm }

\noindent
Fig.2.
The quasinormal-mode frequencies for polar and axial perturbations as
functions of the compactness $ M /R . $ The upper panel shows the
imaginary part and the lower one the real part. The full-drawn curve is
for polar modes and the dashed one for axial modes. The circles
represent the values of the lowest mode of the axial perturbation
obtained by Chandrasekhar  and Ferrari (1991b) and the triangles those
of the higher modes by Kokkotas (1994).  The line indicated by f
represents the frequency of the Kelvin f-mode.

\end{document}